\documentclass[12pt]{article}

\usepackage[letterpaper, margin=1in]{geometry}
\usepackage{float}
\usepackage{amsmath}
\usepackage{booktabs}
\usepackage{multirow}
\usepackage{graphicx}
\usepackage{threeparttable}
\usepackage{wasysym}   
\usepackage[normalem]{ulem}
\usepackage{url}
\usepackage{subcaption}
\usepackage{makecell}  
\usepackage[T1]{fontenc}   
\usepackage{dirtree}  

\usepackage{placeins}

\usepackage[ natbib=true, style=numeric,sorting=none]{biblatex}
\begin{document}

{\Large\bfseries Chapter Title\par}
\vspace{0.5cm}
{\Large \bfseries End-to-End Co-Simulation Testbed for Cybersecurity Research and Development in Intelligent Transportation Systems\par}
\vspace{0.5cm}

{\large\bfseries List of Authors\par}
\textbf{Minhaj Uddin Ahmad}, Ph.D. Student in Transportation Systems Engineering, Department of Civil, Construction, and Environmental Engineering, the University of Alabama, Tuscaloosa, AL, USA

\textbf{Akid Abrar}, Ph.D. Student in Transportation Systems Engineering, Department of Civil, Construction, and Environmental Engineering, the University of Alabama, Tuscaloosa, AL, USA

\textbf{Sagar Dasgupta}, Ph.D., Postdoctoral Fellow at Connected and Automated Laboratory, the University of Alabama, Tuscaloosa, AL, USA

\textbf{Mizanur Rahman}, Ph.D., Assistant Professor in Transportation Systems Engineering, Department of Civil, Construction and Environmental Engineering, the University of Alabama, Tuscaloosa, AL, USA

\vspace{0.5cm}

{\large\bfseries Abstract\par}

Intelligent Transportation Systems (ITS) have been widely deployed across major metropolitan regions worldwide to improve roadway safety, optimize traffic flow, and reduce environmental impacts. These systems integrate advanced sensors, communication networks, and data analytics to enable real-time traffic monitoring, adaptive signal control, and predictive maintenance. However, such integration significantly broadens the ITS attack surface, exposing critical infrastructures to cyber threats that jeopardize safety, data integrity, and operational resilience. Ensuring robust cybersecurity is therefore essential, yet comprehensive vulnerability assessments, threat modeling, and mitigation validations are often cost-prohibitive and time-intensive when applied to large-scale, heterogeneous transportation systems. Simulation platforms offer a cost-effective and repeatable means for cybersecurity evaluation, and the simulation platform should encompass the full range of ITS dimensions—mobility, sensing, networking, and applications. This chapter discusses an integrated co-simulation testbed that links CARLA for 3D environment and sensor modeling, SUMO for microscopic traffic simulation and control, and OMNeT++ for V2X communication simulation. The co-simulation testbed  enables end-to-end experimentation, vulnerability identification, and mitigation benchmarking, providing practical insights for developing secure, efficient, and resilient ITS infrastructures. To illustrate its capabilities, the chapter incorporates a case study on a C-V2X proactive safety alert system enhanced with post-quantum cryptography, highlighting the role of the testbed in advancing secure and resilient ITS infrastructures.
\newpage

\section{Introduction}

Intelligent Transportation Systems (ITS), now deployed in major metropolitan regions across Europe, Asia, and the United States, play a pivotal role in enhancing roadway safety, optimizing traffic flow, and reducing environmental impact. The integration of advanced sensors, communication networks, and data analytics, while enabling real-time traffic monitoring, adaptive signal control, and predictive maintenance, also expands ITS's cyber attack surface, exposing the system to sophisticated cyber threats from malicious actors. One of the foremost concerns in ITS security lies in the security of roadside infrastructure. Components such as traffic signals, sensors, and variable message signs often operate on outdated or poorly maintained systems. Weak login credentials and unpatched software can make them attractive targets for attackers. Denial-of-service and jamming attacks also pose serious challenges~\cite{islam2024review}.

These vulnerabilities are not limited to roadside infrastructure. Inside the vehicle, traditional in-vehicle networks were not designed with cybersecurity as a primary concern. The Controller Area Network (CAN) bus, for example, lacks encryption and authentication, meaning that a compromised node can inject arbitrary messages~\cite{can_bus}. Likewise, physical access points such as the On-Board Diagnostics (OBD) port can provide attackers with direct control over critical vehicle functions if not properly secured~\cite{ammar2020}. As vehicles continue to evolve toward greater connectivity and automation, such weaknesses are no longer confined to local physical attacks but can be exploited remotely once the external network perimeter is breached.

Communication channels introduce yet another dimension of risk. Without robust authentication and encryption, Vehicle-to-Everythng (V2X) messages are vulnerable to interception or spoofing. For example, attackers may exploit unsecured wireless links to inject false information, such as fabricated hazard alerts or manipulated signal phases, that can mislead drivers and compromise traffic safety~\cite{itsa2024future, sae2024j3161}. The reliance on multiple communication technologies, including Cellular V2X (C-V2X), 4G/5G/6G, Wi-Fi, cellular and Bluetooth, further broadens the attack surface, multiplying opportunities for cyber intrusions into both vehicles and infrastructure~\cite{5gaa2019roadmap}. Because ITS applications rely heavily on reliable and time-sensitive wireless communication, flooding the communication spectrum or spoofing Global Navigation Satellite System (GNSS) signals can degrade or even block message delivery. In safety-critical scenarios where split-second decisions matter, the absence or corruption of these time sensitive critical messages can create highly dangerous conditions on the road. In addition, the continuous exchange of data that underpins ITS raises significant privacy concerns. Vehicles routinely transmit sensitive information, including location, identity, and travel patterns. If inadequately protected, this data can be exploited for surveillance, tracking, or manipulation. Beyond endangering individual users, such intrusions risk eroding public confidence in ITS technologies and their broader adoption~\cite{sae2024j3161}.

Implementing robust cybersecurity measures across ITS infrastructures is of utmost importance to ensure continuous Availability, preserve Integrity, and maintain Confidentiality (CIA), the core principles of cybersecurity whose compromise could endanger human safety and operational resilience. The vast scale and heterogeneity of transportation infrastructures and cybersecurity domains, combined with the safety-critical nature of real-world deployments, render comprehensive vulnerability assessments, threat model development, intrusion detection evaluations, and mitigation strategy validations both cost-prohibitive and time-intensive. Simulation platforms provide a safe, repeatable, and cost-effective environment for conducting comprehensive cybersecurity evaluations of ITS infrastructures; however, no single simulator integrates ITS applications, detailed sensor models, realistic mobility patterns, and network connectivity components to represent the full transportation ecosystem. For example, CARs Learning to Act (CARLA)~\cite{Dosovitskiy17} delivers high fidelity 3D world modeling and diverse sensor simulation, Simulation of Urban MObility (SUMO)~\cite{SUMO2018} excels at microscopic traffic dynamics and signal controller modeling, and Objective Modular Network Testbed in C++ (OMNeT++) specializes in detailed network and V2X communication simulations; however, none of these platforms alone captures all critical dimensions - mobility, sensing, network connectivity, and application logic required to emulate a complete ITS ecosystem. The gap is bridged by integrating these individual simulators into an end-to-end co-simulation testbed that couples high-fidelity environment and sensor modeling, comprehensive traffic and signal control simulation, and realistic network and V2X communication simulation to enable rigorous cybersecurity research and development.

This book chapter explains how a co-simulation platform can be constructed and applied to ITS cybersecurity research. The platform comprises (i) CARLA for 3D world modeling and sensor simulation; (ii) SUMO for microscopic traffic simulation and traffic controller modeling; and (iii) OMNeT++ for V2X communication simulation. Through a case study of a C-V2X based proactive safety alert system employing post quantum cryptography (PQC) to harden V2X messaging, and communication domains to achieve cohesive and realistic interactions. Such co-simulation testbeds empower researchers and practitioners to identify cybersecurity vulnerabilities before deployment, benchmark and refine mitigation strategies, and perform rigorous performance evaluations, providing practical guidance for developing secure, efficient, and resilient ITS solutions that meet evolving transportation cybersecurity needs.

\section{Simulation Platforms for ITS}

Simulation is a well-established approach for studying and analyzing complex systems. Instead of experimenting directly in the real world, which is often costly, risky, or infeasible, researchers build simulators to replicate the behaviors of a system in a controlled virtual environment. A simulator creates a model of reality, abstracting away unnecessary details while preserving the dynamics most relevant to the problem at hand. By doing so, it enables safe, repeatable, and large-scale experimentation that would otherwise be impractical. 

The design of a simulator is inherently domain-specific. This is because every complex system, whether physical, social, or technological, has its own set of governing rules, scales, and interactions that must be captured accurately. For example, an aerodynamic simulator must model airflow physics with high precision, while a computer network simulator focuses on packet-level communication protocols. Attempting to build a single universal simulator would be impractical—not only due to the enormous computational demands, but also because of the difficulty of managing, maintaining, and learning such a system. Specialized simulators are therefore created, each optimized to represent a particular domain with the fidelity needed by its users.

At the same time, modern technological systems are increasingly multidisciplinary, cutting across traditional domain boundaries. Transportation, energy, communication, and urban infrastructure, for instance, are deeply interconnected, and innovations in one area often depend on interactions with others. As the world becomes more cross-domain and multidisciplinary, the limitations of standalone simulators become evident. To address this, co-simulation techniques are employed, where multiple specialized simulators are executed together and exchange information in real time. This approach allows researchers to study complex systems-of-systems while leveraging the strengths of each simulator without overburdening any single tool.

In the context of ITS, simulation platforms are indispensable for design, testing, and validation. Modern ITS applications extend well beyond traditional traffic management, incorporating connected and autonomous vehicles, real-time data exchange, and smart infrastructure. At a minimum, effective ITS applications involve three complementary domains.

\begin{itemize}
    \item \textbf{Vehicle dynamics and traffic flow modeling}, to capture both individual vehicle behavior and large-scale traffic interactions across roadway networks.
    
    \item \textbf{Sensor and environment modeling}, to reproduce how vehicles perceive their surroundings through cameras, LiDAR, radar, and other sensing modalities.
    
    \item \textbf{Connectivity and communications modeling}, to simulate vehicle-to-vehicle (V2V), vehicle-to-infrastructure (V2I), and broader C-V2X communication that enable cooperative and coordinated mobility.
\end{itemize}

No single simulator can natively cover all of these domains with sufficient detail. As a result, researchers rely on specialized simulation platforms and, increasingly, on co-simulation frameworks that integrate multiple simulators into a unified experimental environment. The following sections provide an overview of these specialized standalone simulators, co-simulation platforms and their roles in ITS research.

\subsection{Standalone Specialized Simulators}

Building on the three domains introduced earlier, standalone specialized simulators for ITS can be grouped into (i) vehicle dynamics and traffic flow simulators, (ii) sensor and environment modeling simulators, and (iii) connectivity and communication simulators. Each group addresses a distinct aspect of modern ITS, offering depth within its scope but remaining limited outside of it.  

Vehicle dynamics and traffic flow simulators such as SUMO, PTV Vissim~\cite{ptv_vissim}, and Aimsun~\cite{aimsun2025} model large-scale roadway traffic. Well-established commercial tools such as PTV Vissim and Aimsun are widely used in professional traffic engineering practice. They provide mature features, strong visualization capabilities, and extensive support for roadway design tools and operations analysis, making them valuable for consultants, practitioners, and government agencies. However, because they are closed-source, these platforms offer limited flexibility for experimental modifications and are less suited for exploratory research. In contrast, SUMO has emerged as the preferred choice in academic and research settings. In addition to offering many of the core features found in these commercial platforms, SUMO has emerged as the preferred choice in academic and research settings. As a highly scalable, open-source microscopic simulator, it can model thousands of vehicles across extensive roadway networks. SUMO implements well-known car-following models such as the Intelligent Driver Model (IDM) and Krauss, along with lane-changing models including DK2008, LC2013, and SL2015, enabling detailed representation of vehicle interactions and lane-level dynamics~\cite{sumoSL2015, SUMO2018, erdmann2015sumoLC2013}. It also supports route assignment, multimodal traffic, and traffic signal control. Most importantly, SUMO’s openness and extensibility allow researchers to adapt models, extend functionalities, and integrate with external tools via interfaces such as TraCI~\cite{wegener2008traci}. These features make it especially attractive for academia and cutting-edge ITS research, where flexibility and customizability are often more critical than polished user interfaces.

\textbf{Sensor and environment modeling simulators} focus on replicating how sensors perceive and interact with their surroundings, whether mounted on vehicles or deployed as part of smart infrastructure. Their distinctive contribution lies in providing physics-based simulation of sensing modalities such as LiDAR, cameras, radar, and GNSS, within photorealistic 3D virtual environments. By leveraging modern game engines like Unreal Engine and Unity, these platforms simulate not only the geometry of the environment but also its physical properties, including lighting, weather, and material reflectivity. This enables realistic generation of sensor data streams for testing perception algorithms in autonomous vehicles as well as roadside monitoring systems, supporting sensing applications in both autonomy and intelligent infrastructure. Prominent examples include CARLA, LGSVL~\cite{rong2020lgsvl}, and AirSim~\cite{airsim2017fsr}. CARLA has become the de facto open-source standard among these simulators, offering configurable environmental conditions, synchronous and deterministic time-stepping that enables precise integration with other simulators. LGSVL was originally designed to support Apollo~\cite{apollo2025} and Autoware~\cite{autoware2025} and provided similar features, though its development was suspended in 2022. AirSim, created by Microsoft Research, was initially built for drones and later extended to ground vehicles; however, its support for vehicle-related applications remained limited, and development was discontinued in 2023. These simulators are also valuable in reinforcement learning, AI, and perception research because they allow safe and repeatable experimentation with complex sensing setups. At the same time, their scope is narrower than that of traffic simulators: they do not model large-scale flows using microscopic traffic models, nor do they natively incorporate V2X communication stacks. This limits their applicability when evaluating connected transportation systems in a holistic way.

\textbf{Connectivity and communication simulators} provide the third domain, focusing on networking and wireless communication. Tools such as OMNeT++~\cite{omnetpp} and ns-3~\cite{ns3website} were originally developed as general-purpose computer network simulators and include comprehensive support for internet protocol stacks, from traditional IP and transport protocols to advanced wireless standards. Their flexibility and extensibility have made them the foundation for modeling vehicular communication technologies as well. Building on their core networking capabilities, researchers have introduced support for vehicular communication standards, including DSRC, LTE, and emerging C-V2X technologies. Extensions such as Veins~\cite{veins2025} add protocol-level implementations of V2X communication, while other frameworks adapt OMNeT++ and ns-3 for studying wireless channel behavior and cooperative applications in transportation systems. Between the two, OMNeT++ is often favored in ITS contexts because of its modular design, visual debugging tools, and the ease with which it can be extended. ns-3, by contrast, provides lower-level and highly detailed protocol modeling, offering greater fidelity at the cost of accessibility and visualization. Both simulators remain central in ITS research as both bring the breadth of computer network simulation into the transportation domain, enabling exploration of how connectivity interacts with mobility and autonomy.

These specialized simulators illustrate how each of the three domains has its own dedicated tools that excel within their boundaries. While indispensable, they cannot individually capture the cross-domain interactions that characterize modern ITS, motivating the use of co-simulation approaches discussed in the next subsection.

\subsection{Co-Simulation of Multiple Specialized Simulators}
As discussed earlier, standalone simulators excel within their domains-traffic flow, sensor and environment modeling, or networking, but remain limited for comprehensive ITS applications when applied in isolation. Traffic simulators such as SUMO or Vissim capture large-scale mobility patterns yet lack perception and connectivity. Driving simulators like CARLA provide photorealistic environments and sensor physics but are decoupled from standardized microscopic traffic models. Network simulators such as OMNeT++ or ns-3 support rich protocol stacks, but their mobility models are typically abstract and fail to reflect realistic vehicle dynamics.

These siloed approaches restrict the ability to evaluate closed-loop interactions among mobility, perception, decision-making, and connectivity. Such interactions are essential in scenarios involving mixed autonomy, cooperative driving, or cybersecurity in transportation applications, where the interplay between sensing, planning, and communication directly shapes outcomes. The absence of modular and extensible integration also hampers experimentation with new algorithms, protocols, or coordination strategies. Co-simulation addresses this gap by bridging specialized simulators into unified frameworks, enabling multi-domain studies that none of the tools can support individually.

One example is the native SUMO–CARLA co-simulation, introduced in CARLA v0.9.8, which synchronizes CARLA’s high-fidelity vehicles and sensor models with SUMO's microscopic traffic simulation via an extended TraCI interface. This integration allows hybrid scenarios where CARLA manages ego vehicles with detailed dynamics and perception, while SUMO controls surrounding traffic and infrastructure logic. Several works have extended this baseline: Li et al. developed a synchronized framework for multi-vehicle AV testing, Cantas and Guvenc designed a reinforcement learning–driven traffic co-simulation environment, and Roccotelli et al.~\cite{cantas2023customized, roccotelli2024co} proposed a modular CARLA–SUMO platform for studying mixed traffic autonomy. While these efforts bridge traffic modeling and perception, most lack C-V2X support, relying on simplified abstractions for V2V communication.

In the networking domain, OMNeT++ and ns-3 have been extended to include vehicular communication, leading to frameworks such as Veins, OpenCV2X~\cite{mccarthy2021opencv2x} and others. Among these, Veins has become the most widely used for ITS, offering a robust SUMO–OMNeT++ co-simulation pipeline. Veins enables bidirectional synchronization of traffic mobility and communication protocols, including DSRC, and even supports channel attenuation modeling with building polygons. However, its vehicles are limited to SUMO's node abstractions, and it lacks high-fidelity sensor modeling. For sensing and perception algorithm development and testing, platforms like CARLA remain necessary. 

To overcome these limitations, newer co-simulation frameworks extend the scope further. Veins-Carla~\cite{hardes2023poster} replaces SUMO with CARLA to combine detailed vehicle dynamics with communication modeling. MS-VAN3T-CARLA~\cite{carletti2024ms} connects CARLA and ns-3 using gRPC middleware, enabling photorealistic cooperative perception experiments with realistic C-V2X communication. CARLA–SUMO–Artery~\cite{bouchouia2022simulator} integrates the Artery stack to support ETSI ITS-G5 protocols for security and cooperative automation studies. OpenCDA~\cite{xu2021opencda} provides a unified platform for cooperative maneuvers such as platooning and coordinated merging. SimuTack~\cite{simutack:2023} integrates SUMO, CARLA, and OMNeT++ into a single co-simulation pipeline, while Eclipse MOSAIC acts as middleware, providing modular adapters to couple multiple domain-specific simulators. MOSAIC~\cite{schrab2022MOSAIC} functions as a modular co-simulation middleware connecting domain-specific simulators through dedicated adapters for each simulator. While offering flexible and standardized interface between individual simulators, the CARLA adapter is still under development. MOSAIC Extended, a commercial version of MOSAIC provides support of 3D environment and vehicle physics simulation through a proprietary simulator PHABMACS~\cite{phabmacs}. However, PHABMACS restricts environmental realism by procedurally generating minimal road furniture (no curbs, lamp posts, or benches), and targets small-scale ADAS prototyping trading off high-detail physics for fewer vehicles support.

\begin{table}[!htbp]
\centering
\footnotesize

\newcommand*\feature[1]{\ifcase#1 \Circle\or\LEFTcircle\or\CIRCLE\fi}
\newcommand*\f[4]{\feature#1 & \feature#2 & \feature#3 & \feature#4}

\begin{threeparttable}
\caption{Comparison of co-simulation platforms}
\label{tab:comparison-table}
\begin{tabular}{@{}l cccc@{}}
\toprule
Platform  & \makecell{Environment \& \\ Perception Sensor} & \makecell{Traffic \\ Simulation} & \makecell{Communication \\ Simulation} & Generalization \\
\midrule

Veins~\cite{sommer2011bidirectionally} & \f0220 \\
Veins-Carla~\cite{hardes2023poster} & \f2120 \\
ms-van3t-carla~\cite{carletti2024ms} & \f2121 \\
Carla-SUMO-Artery~\cite{bouchouia2022simulator} & \f2220 \\
PLEXE~\cite{segata2022multi} & \f0221 \\
OpenCDA~\cite{xu2021opencda} & \f2211 \\
Simutack~\cite{simutack:2023} & \f2220 \\
MOSAIC~\cite{schrab2022MOSAIC} & \f1222 \\
\textbf{OpenCAMS} & \f2222\\

\bottomrule
\end{tabular}
\begin{tablenotes}[para, flushleft]
\item $\feature2=\text{full support}$
\item $\feature1 = \text{partial support}$ 
\item $\text{\feature0}=\text{no support}$
\end{tablenotes}
\end{threeparttable}
\end{table}

Table~\ref{tab:comparison-table} summarizes these co-simulation platforms against three essential capabilities for Connected and Automated Mobility (CAM): high-fidelity environment and perception modeling, scalable microscopic traffic simulation, and flexible C-V2X communication support. The comparison highlights remaining research gaps. Many frameworks support synchronous co-simulation but are narrowly tailored to specific application domains. Communication stacks are often tightly coupled to particular protocols, limiting generalizability. Moreover, synchronization mechanisms may not be modular, making it challenging to extend these platforms to diverse scenarios or additional simulation modules. These gaps underline the need for flexible, modular, and open co-simulation frameworks that can evolve alongside cutting-edge ITS research.

\section{Co-Simulation Testbed for ITS Cybersecurity}

To address these critical challenges, Ahmad et. al.~\cite{ahmad2025opencams} a generalized and time-synchronized co-simulation platform OpenCAMS that integrates three unique simulators for Connected and Autonomous Mobility (CAM) applications in ITS: (i) SUMO for scalable traffic modeling, (ii) CARLA for high-fidelity environmental perception and vehicle dynamics, and (iii) OMNeT++ for flexible C-V2X communication simulation. The co-simulation platform, OpenCAMS enables researchers to explore the interplay between large-scale traffic dynamics, detailed vehicle perception, and real-time C-V2X communications within a single, cohesive environment. Beyond cybersecurity research, by allowing the simulation of complex cyberattack scenarios and mitigation approaches such as signal spoofing, message tampering, and denial-of-service attacks across all layers of the transportation ecosystem, OpenCAMS also offers significant potential for improving safety and mobility. Its extensible design supports the integration of emerging technologies and threat models, making it a valuable tool for developing and testing robust mitigation strategies.

\begin{figure}[!htbp]
    \centering
    \includegraphics[width=0.8\linewidth]{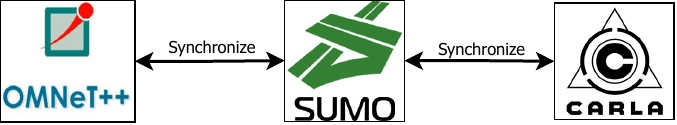}
    \caption{Simulator components of OpenCAMS}
    \label{fig:cosim-logos}
\end{figure}

Figure~\ref{fig:cosim-logos} present the indiviual simulator components of OpenCAMS co-simulation platform. OpenCAMS employs a multi-client TraCI loop with strict time-step synchronization to ensure consistency across all three simulators. The unique feature of this co-simulation platform is that it avoids adding any abstraction layer on top of the simulators, preserving the flexibility and full potential of the underlying simulators. CARLA may simulate only a subset of vehicles requiring advanced in-vehicle sensors or roadside sensors, while SUMO handles the background roadway traffic. OMNeT++ dynamically maps communication nodes to vehicles and infrastructures, leveraging wireless propagation models and customizable communication networking stacks. This design supports reproducible, modular, and extensible research across various domains, including the development and evaluation of reactive and proactive safety, mobility, and cybersecurity integrations related to CAM applications. In essence, our platform enables a unified, scalable, and realistic testbed for transportation research that bridges the gap between what each simulator is capable of in isolation and what modern connected and automated transportation systems demand.

The development of OpenCAMS, as detailed in ~\cite{ahmad2025opencams}, was driven by the critical need to address the limitations of existing simulation platforms in capturing the full complexity of modern ITS systems, prioritizing open-source components, minimum abstraction and support of generalized ITS applications. By fostering a collaborative, accessible environment, OpenCAMS can empower research community to advance secure, efficient, and resilient next-generation transportation systems. For the latest details on its implementation, readers are encouraged to consult the project’s GitHub repository at https://github.com/minhaj6/carla-sumo-omnetpp-cosim.

The OpenCAMS co-simulation platform architecture enables integrated development, evaluation, and validation ensuring coherent evolution across cyber-physical layers. Diverse application domain includes safety, mobility, and cybersecurity.

\textbf{Safety:} OpenCAMS allows rigorous testing of safety-critical applications such as Forward Collision Warning, Blind Spot Warning/Lane Change Assist, and Vulnerable Road User protection. CARLA models suit of sensors, OMNeT++ transmits Basic Safety Messages (BSMs) and cooperative perception messages, while SUMO manages flow of traffic in the road network. Researchers can test robustness under varying weather, visibility, latency, and traffic densities.

\textbf{Mobility:} The platform supports platooning, cooperative merging, adaptive following, and smart corridor management. SUMO manages background traffic and signals, CARLA handles decision logic and perception, and OMNeT++ simulates CAM, V2I, and V2N communication. Applications like Green Light Optimal Speed Advisory (GLOSA), energy-aware routing, fleet electrification, and emergency vehicle preemption can be studied with end-to-end impacts on throughput, fuel economy, and safety.

\textbf{Cybersecurity:} OpenCAMS provides a controlled environment for evaluating attacks such as Sybil, replay, and infrastructure manipulation. OMNeT++ application layer can model malicious messages, CARLA can model target vehicle responses, and SUMO evaluates network-wide consequences. Researchers can test detection and mitigation mechanisms including authentication, timestamp verification, and anomaly detection.

OpenCAMS allows for comprehensive ITS cybersecurity research by enabling the simulation of cyber attacks in a controlled environment. For example, researchers can simulate Sybil attacks, where multiple fake identities are injected into the network; replay attacks, where valid messages are retransmitted maliciously; or infrastructure compromise, where traffic signals are manipulated. This integrated approach allows for the evaluation of vulnerabilities across the entire system, from perception to communication, and the testing of mitigation strategies like intrusion detection systems or cryptographic enhancements.  Overall, OpenCAMS provides a closed-loop, scenario-based benchmarking platform for emerging research in cooperative safety, mobility optimization, cybersecurity, and digital twin systems. Its extensibility supports frontiers such as human-autonomy interaction, teleoperation, and city-scale integration. 

The following section presents an example implementation of an infrastructure-enabled crash warning system that uses roadside sensors to detect vehicles approaching an urban intersection, calculates Time-to-Collision (TTC) between as they approach the intersection, and broadcasts warning messages verified by Post Quantum Cryptography (PQC) signatures.

\section{Case Study: Demonstration of OpenCAMS for ITS Cybersecurity Research}

Central to the operation and effectiveness of modern ITS architectures is connectivity. Connectivity through C-V2X enables a wide range of ITS applications that enhance security, safety, mobility, and operational efficiency. For instance, queue warning systems notify vehicles about slow or stopped traffic ahead, helping drivers react in time and reducing the likelihood of rear-end collisions. Emergency vehicle alerts are another critical use case where drivers and connected systems get alerts about the approach and intended route of ambulances, fire trucks, or police vehicles, enabling surrounding traffic to clear lanes in advance~\cite{sae2024j3161/1}. Cooperative Adaptive Cruise Control (CACC) further demonstrates the potential of V2V communication by synchronizing acceleration and braking across multiple vehicles, which supports safe platooning, improves fuel efficiency, and helps ease congestion~\cite{liu2021survey, 5gaa2019roadmap}. Another prominent example of a safety application is the Intersection Collision Avoidance (ICA) warning system, where vehicles approaching a connected intersection receive broadcasts from a roadside unit (RSU) if the smart intersection detects any potential conflicts, improving intersection safety~\cite{sae2016j2735, sae2024j3161}.

C-V2X communication enables real-time information sharing between vehicles, infrastructure, and cloud services. C-V2X supports both PC5 sidelink for direct, low-latency communication and Uu interface over 4G/5G networks for broader coverage and cloud integration. These capabilities are indispensable for latency-sensitive and cooperative applications such as platoon control, intersection coordination, cooperative perception, and dynamic traffic signal control. Today, connectivity is no longer optional; it is foundational. It enables not just functional capabilities but also determines the system’s adaptability and resilience. However, this reliance on connectivity further enlarges the attack surface for adversaries, who may exploit vulnerabilities at the physical, communication, or application layers. Consequently, testing the cyber resilience of connected transportation systems has become an urgent research priority. Evaluating these systems under real-world cyber-physical attack scenarios is critical but fraught with logistical, financial, and regulatory challenges.

\begin{figure}[!htbp]
    \centering
    \includegraphics[width=\linewidth]{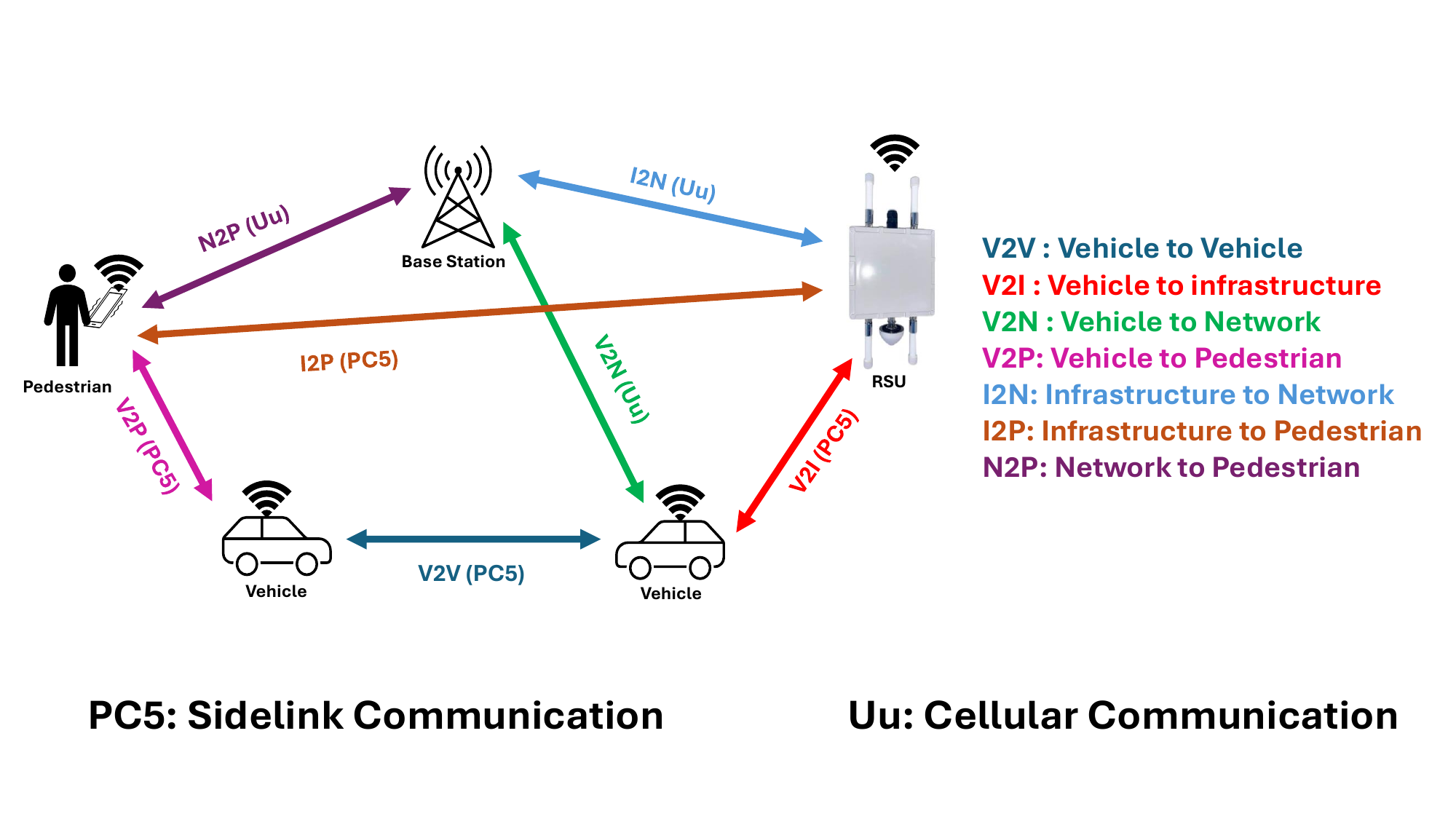}
    \caption{Communication modes of C-V2X: direct PC5 sidelink for V2V, V2I, I2P, and V2P, and Uu interface for V2N, N2P, and I2N}
    \label{fig:cv2x_overview}
\end{figure}

C-V2X is a communication technology standardized by 3GPP starting in Release 14~\cite{3gpp2017rel14}. It was designed to support direct communication between vehicles, infrastructure, and vulnerable road users as well as network-based communication through cellular operators~\cite{sunuwar2024crosslayer}. These communication capabilities form the backbone for a wide range of ITS applications. C-V2X includes two complementary modes. The first is the PC5 sidelink mode, which enables direct V2V, V2I, and V2P communications over the 5.9 GHz ITS band without relying on cellular coverage~\cite{molina2017mode4}. The second is the Uu interface, which provides vehicle-to-network (V2N) communication using LTE or 5G infrastructure \cite{5gaa2019roadmap}. Together, these modes allow both low-latency safety message exchanges with surrounding vehicles and roadside infrastructure, as well as broader coverage through the cellular network, including communication with cloud services. Figure~\ref{fig:cv2x_overview} presents the communication modes of C-V2X.

For the safety-critical sidelink connectivity, the 3GPP release 14 introduced two new communications standards, Mode 3 or scheduled mode, and Mode 4 or autonomous mode. In Mode 3, the cellular infrastructure assists in scheduling communication resources for device-to-device, which requires the devices to be inside the cellular coverage area. However, in Mode 4, vehicles can autonomously select and manage their radio resources using sensing-based Semi-Persistent Scheduling (SPS). Mode 4 is especially important because safety applications must operate even where no cellular coverage is available. Subsequent releases, such as Release 16, extended C-V2X into 5G New Radio V2X (NR-V2X), improving reliability and adding support for advanced use cases like cooperative perception and high-density platooning \cite{sae2024j3161}.

Intersection Collision Avoidance is recognized as one of the most critical safety applications in ITS. Intersections are complex and high-risk environments where vehicles, pedestrians, and signals converge. A significant share of crashes and traffic fatalities occur at intersections, and many involve angle collisions that are particularly severe. ICA applications provide early warnings to approaching vehicles, helping drivers or automated systems to react in time and avoid dangerous conflicts. This ability to prevent side-impact collisions makes ICA a cornerstone of safety-critical ITS services~\cite{itsa2024future}.  

\begin{figure}[!htbp]
    \centering
    \includegraphics[width=\linewidth]{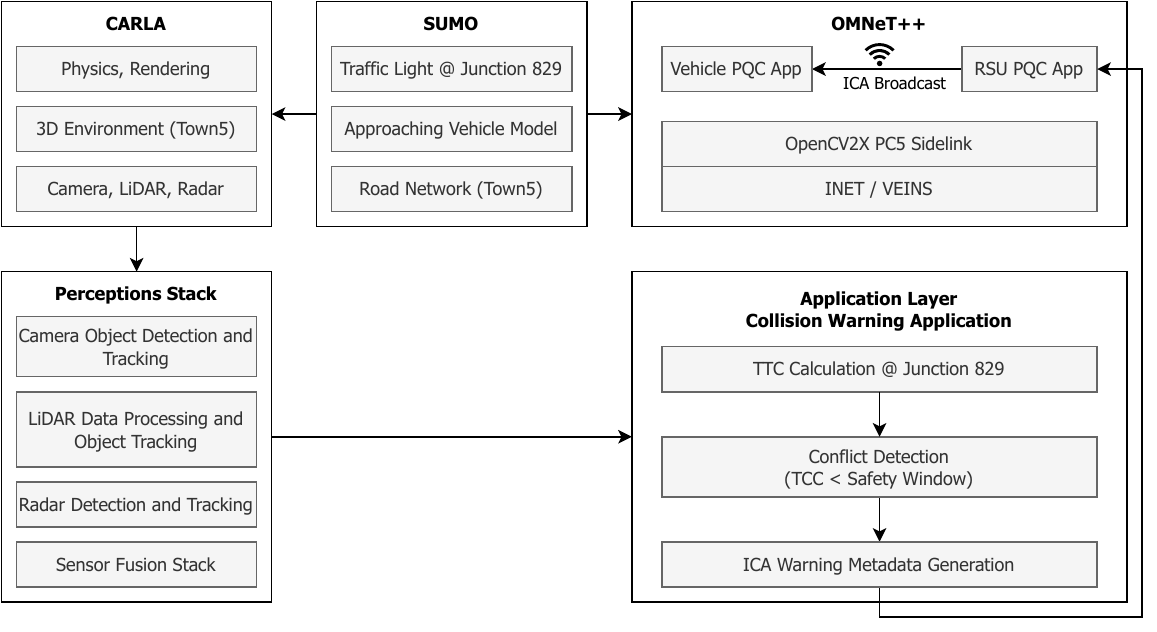}
    \caption{Functional diagram of PQC evaluation in OpenCAMS: TTC at junction 829 and ICA broadcast}
    \label{fig:cosim-application}
\end{figure}

While ICA requires reliable traffic modeling and communication, it also depends on the trustworthiness of the warning messages themselves. PQC provides digital signatures that remain secure and trustworthy against both traditional cyber attacks and in the presence of future quantum adversaries that could break classical algorithms such as Rivest–Shamir–Adleman (RSA) or Elliptic Curve Digital Signature Algorithm (ECDSA)~\cite{nist2022pqc,chen2016report}. By integrating PQC into V2X communication, ITS applications ensure that any C-V2X message exchange, including ICA warnings cannot be forged or tampered with. Falcon, a lattice-based signature algorithm standardized by NIST, is particularly well-suited for connected applications. It produces compact signatures and offers very fast verification, both of which are essential for high-volume, low-latency V2X messaging~\cite{nist2022pqc,falcon2019spec,giuliari2023pqc}. Evaluating PQC in a co-simulation environment is therefore required to confirm that stronger security does not undermine real-time safety. In this work, Falcon is adopted for ICA, providing both quantum-resistant security and practical performance for time-critical transportation safety applications.Conducting such evaluations in the real world is neither economical nor safe due to the risks of traffic disruption and potential hazards. This makes the OpenCAMS co-simulation testbed a vital tool for advancing secure and resilient ITS applications. Figure~\ref{fig:cosim-application} presents the functional diagram of PQC enabled ICA application in OpenCAMS platform. 

\subsection{Implementation of PQC in ICA Application}

We selected a signalized intersection in CARLA Town 05 (Junction 829). We created a SUMO route file that produces eight vehicles that pass through the chosen intersection. We placed a roadside unit (RSU) in OMNeT++ at the side of the intersection. We then spawned the same vehicles in CARLA using the SUMO routes so that traffic, sensing, and communication are aligned in time. We built a sensor view from the roadside unit's point of view. The RSU is equipped with LiDAR and radar. We defined the geometric center of the intersection as the reference point for calculating TTC. In every SUMO simulation step, we computed the time for each vehicle to reach the reference point from its current location. If the time difference between any two vehicles was smaller than two seconds, the roadside unit generated an ICA message. For the message structure, we followed the SAE J2735 standard~\cite{sae2016j2735}, specifically the \texttt{MSG\_IntersectionCollisionAvoidance} format, when defining and transmitting the ICA messages within OMNeT++. In addition, we adopted an X.509 certificate style to attach certificates directly to the ICA messages. The certificates carry the verification keys, enabling the receiving vehicles to validate the authenticity of the broadcast message. Both vehicles and the roadside unit use C-V2X PC5 Mode 4 links for direct communication. As soon as an ICA message is created, the roadside unit signs the message with the Falcon-512 post-quantum digital signature. The roadside unit also builds a certificate that includes the public verification key and broadcasts the signed warning message.

\begin{figure}[!htbp]
    \centering
    \begin{subfigure}[b]{0.48\textwidth}
        \centering
        \includegraphics[width=\textwidth]{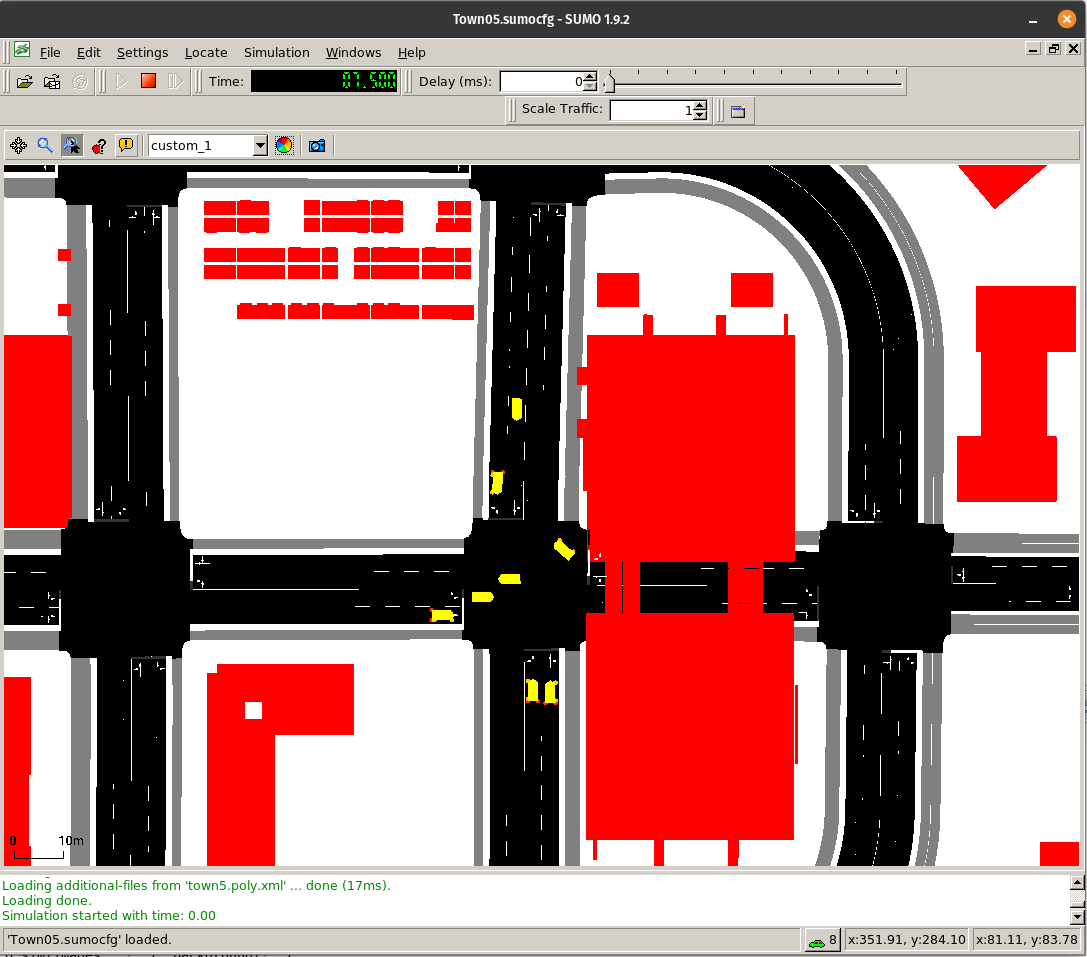}
        \caption{SUMO interface in co-simulation}
        \label{fig:sumo-map}
    \end{subfigure}
    \hfill
    \begin{subfigure}[b]{0.48\textwidth}
        \centering
        \includegraphics[width=\textwidth]{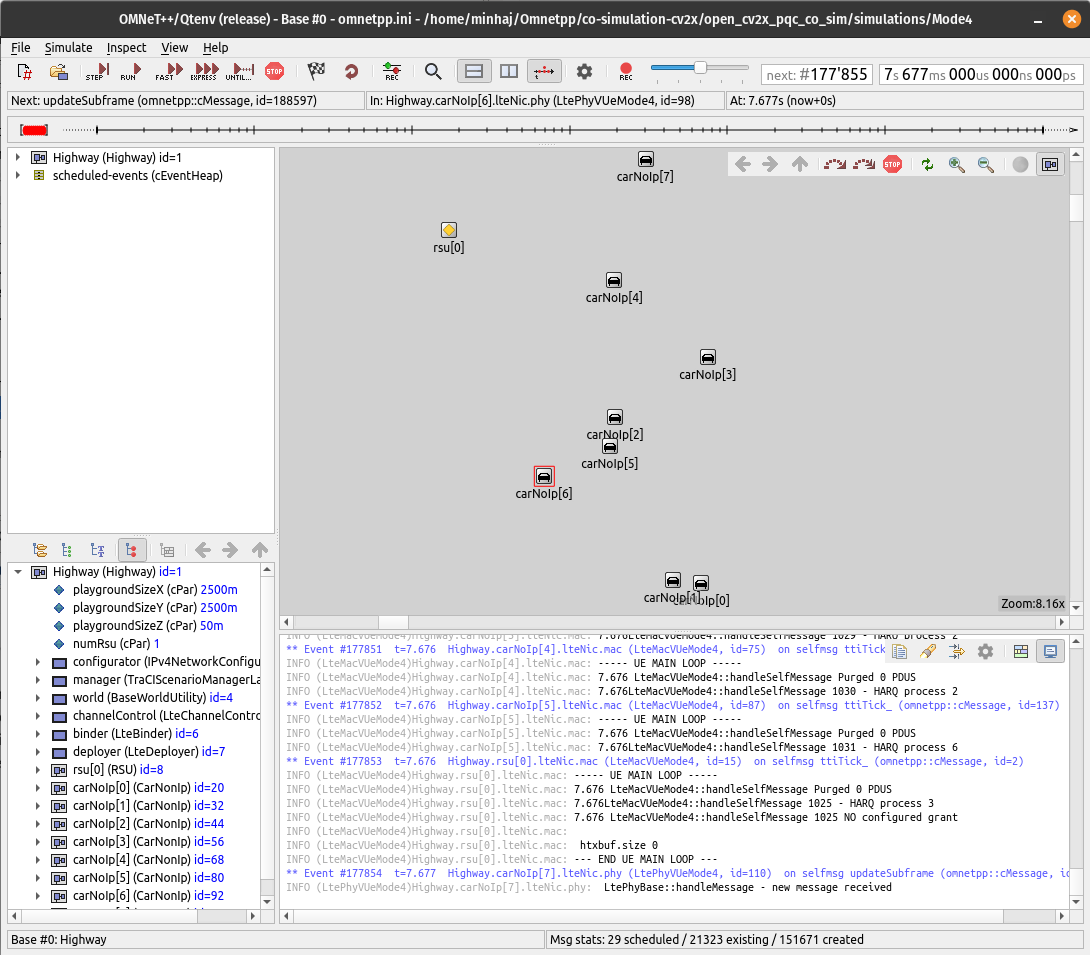}
        \caption{OMNeT++ interface in co-simulation}
        \label{fig:omnet-map}
    \end{subfigure}

    \begin{subfigure}[b]{0.48\textwidth}
        \centering
        \includegraphics[width=\textwidth]{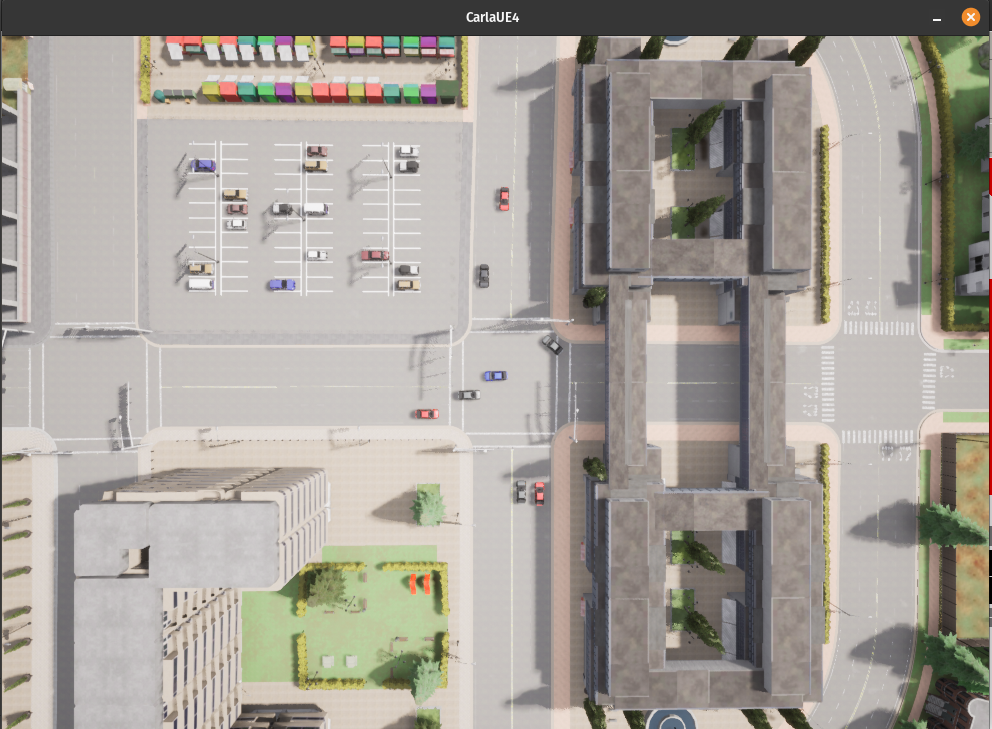}
        \caption{CARLA bird’s eye view}
        \label{fig:carla-bev}
    \end{subfigure}
    \hfill
    \begin{subfigure}[b]{0.48\textwidth}
        \centering
        \includegraphics[width=\textwidth]{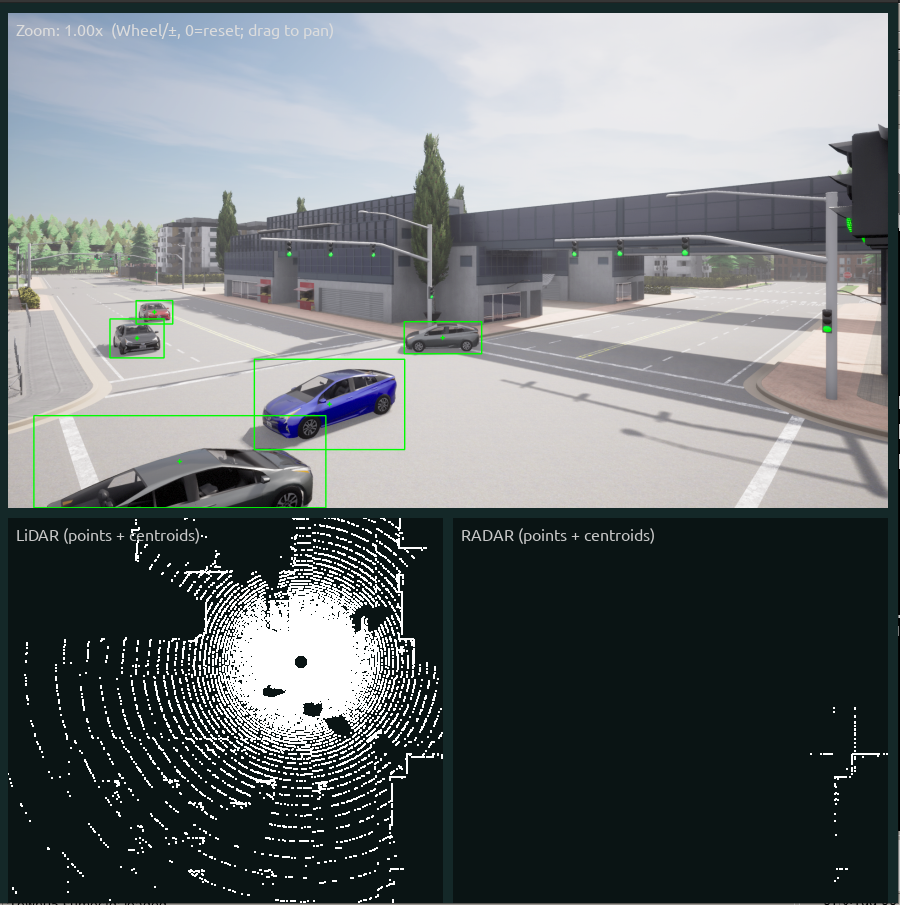}
        \caption{CARLA perception view from RSU}
        \label{fig:carla-perception}
    \end{subfigure}

    \caption{SUMO, OMNeT++, and CARLA running synchronously in the OpenCAMS co-simulation.}
    \label{fig:all-sim}
\end{figure}

Figure~\ref{fig:all-sim} shows the integrated view of the three simulators in the OpenCAMS co-simulation platform. The top row shows the (a) SUMO and (b) OMNeT++ interfaces. The bottom row shows the (c) CARLA bird’s eye view and the (d) roadside perception view. This four-panel figure illustrates the time-synchronized co-simulation where mobility, sensing, and communication are active at the same timestamp.

\FloatBarrier

\subsection{Metrics and Evaluation of PQC-enabled ICA}

We evaluated the warning pipeline end-to-end using metrics that capture both cryptographic performance and communication reliability. These metrics are critical for verifying that warning messages remain both trustworthy and effective in traffic.

\textbf{Signature generation and verification time:}  
The cryptographic performance is measured in terms of signature generation time at the roadside unit and verification time at the vehicles. If $t_{gen}$ is the average signing time and $t_{ver}$ is the average verification time, then
\[
t_{gen} = \frac{1}{N}\sum_{i=1}^{N} t_{gen}^{(i)}, \quad 
t_{ver} = \frac{1}{N}\sum_{i=1}^{N} t_{ver}^{(i)}
\]
where $N$ is the number of messages signed or verified.

Let $V$ be the number of vehicles. Let $N_v$ be the number of signatures verified by vehicle $v$. Let $t_{\mathrm{ver},v}^{(i)}$ be the $i$th verification time at vehicle $v$. The pooled mean verification time across all vehicles is
\[
\bar t_{\mathrm{ver}}=\frac{\sum_{v=1}^{V}\sum_{i=1}^{N_v} t_{\mathrm{ver},v}^{(i)}}{\sum_{v=1}^{V} N_v}.
\]
The pooled standard deviation is
\[
s_{\mathrm{ver}}=\sqrt{\frac{\sum_{v=1}^{V}\sum_{i=1}^{N_v} \left(t_{\mathrm{ver},v}^{(i)}-\bar t_{\mathrm{ver}}\right)^2}{\sum_{v=1}^{V}N_v-1}}.
\]
For signing at the RSU, if $N_{\mathrm{sig}}$ warnings were signed, the mean signing time is
\[
\bar t_{\mathrm{gen}}=\frac{1}{N_{\mathrm{sig}}}\sum_{i=1}^{N_{\mathrm{sig}}} t_{\mathrm{gen}}^{(i)}.
\]

\textbf{Packet Delivery Ratio (PDR):}  
PDR represents the reliability of message delivery over the PC5 link. It is defined as
\[
PDR = \frac{N_{recv}}{N_{exp}}
\]
where $N_{recv}$ is the number of ICA messages successfully received by vehicles, and $N_{exp}$ is the total number of messages expected.

\textbf{Channel Busy Ratio (CBR):}  
CBR measures the fraction of time the wireless channel is occupied. If $T_{busy}$ is the total time the channel is sensed as busy and $T_{total}$ is the total observation time, then
\[
CBR = \frac{T_{busy}}{T_{total}}.
\]
Monitoring CBR is important because larger post-quantum signatures could increase network load.


We measured signature generation at the RSU and signature verification at the vehicle using the Falcon-512 post-quantum scheme. On average, the roadside unit required 0.30 ms to generate a signature, with a standard deviation of 0.09 ms. The vehicles required an average of 0.11 ms to verify each signature, with a standard deviation of 0.02 ms.

The measured PDR was 0.93 for all the simulated vehicles, as shown in Table~\ref{tab:pdr}, meaning that 93\% of the broadcasted warning messages were successfully received. A high PDR ensures that vehicles consistently receive safety-critical messages, which is vital for preventing collisions even when cryptographic overhead is present.

\begin{table}[!htbp]
\centering
\caption{Summary of ICA message delivery reliability}
\label{tab:pdr}
\begin{tabular}{|c|c|c|c|}
\hline
Vehicle ID & ICA messages expected & ICA messages received & PDR \\
\hline
1 & 61 & 57 & 0.93 \\
2 & 61 & 57 & 0.93 \\
3 & 61 & 57 & 0.93 \\
4 & 61 & 57 & 0.93 \\
5 & 61 & 57 & 0.93 \\
6 & 61 & 57 & 0.93 \\
7 & 61 & 57 & 0.93 \\
8 & 61 & 57 & 0.93 \\
\hline
\end{tabular}
\end{table}

The mean CBR was approximately 0.02 with a maximum standard deviation of 0.007, as shown in Table~\ref{tab:cbr}. This low value indicates that PQC signatures did not create congestion, and there remained sufficient capacity for other V2X messages.

\begin{table}[!htbp]
\centering
\caption{Channel Busy Ratio statistics across vehicles}
\label{tab:cbr}
\begin{tabular}{|c|c|c|}
\hline
Vehicle ID & CBR mean & CBR standard deviation \\
\hline
1 & 0.020 & 0.007 \\
2 & 0.021 & 0.006 \\
3 & 0.023 & 0.003 \\
4 & 0.024 & 0.004 \\
5 & 0.021 & 0.006 \\
6 & 0.020 & 0.007 \\
7 & 0.022 & 0.005 \\
8 & 0.023 & 0.005 \\
\hline
\end{tabular}
\end{table}

Table~\ref{tab:summary_metrics} summarizes the main metrics used in our evaluation. Signature processing times quantify the computational cost of PQC, PDR shows communication reliability, and CBR reflects channel utilization. Together, these metrics provide a balanced view of cryptographic and network performance, ensuring that stronger security does not undermine the usefulness of ICA safety applications.

\begin{table}[!htbp]
\centering
\caption{Application-level metrics for ICA warnings}
\label{tab:summary_metrics}
    \begin{tabular}{|p{4cm}|p{4cm}|p{6cm}|}
    \hline
    \textbf{Metric} & \textbf{Measured value} & \textbf{Relevance for security assessment} \\
    \hline
    Signature generation time & 0.30 ms (mean), 0.09 ms (std. dev.) & Demonstrates whether the RSU can generate signatures for warning messages within acceptable time constraints. \\
    \hline
    Signature verification time & 0.11 ms (mean), 0.02 ms (std. dev.) & Demonstrates whether vehicles can verify incoming warnings promptly to maintain real-time responsiveness. \\
    \hline
    Packet Delivery Ratio & 0.93 & Reflects the reliability of ICA message dissemination in the presence of PQC operations. \\
    \hline
    Channel Busy Ratio & 0.02 (mean), 0.007 (std. dev.) & Indicates the extent of wireless channel utilization and whether PQC signatures contribute to potential congestion. \\
    \hline
    \end{tabular}
\end{table}

\FloatBarrier

\section{Summary}
We discussed ITS-specialized simulators and co-simulation platforms. In addition, we presented a case study using OpenCAMS, a co-simulation testbed that integrates SUMO, CARLA, and OMNeT++ to model CAM systems for transportation cybersecurity research. The utility of co-simulation testbed was further demonstrated through the implementation of a PQC-enabled safety application. Looking ahead, co-simulation platforms with capabilities like those of OpenCAMS can support emerging technologies and next generation ITS research. For instance, integrating 5G NR sidelink protocols, ROS2-based autonomy stacks, real-time GNSS spoofing models, and cyberattacks on cooperative driving. As CAM systems continue to advance, co-simulation remains a critical tool for exploring safe, intelligent, efficient, and secure mobility solutions under realistic and tightly integrated simulation conditions.

\printbibliography 
\end{document}